\documentclass[useAMS,usenatbib]{mn2e}

\usepackage{graphicx,amssymb,color}
\usepackage[normalem]{ulem}

\title[Eclipse, transit, and occultation at exo-syzygy]
{Eclipse, transit and occultation geometry of planetary systems at exo-syzygy}

\author[Veras \& Breedt]{
Dimitri Veras$^{1}$\thanks{E-mail: d.veras@warwick.ac.uk},
Elm\'{e} Breedt$^{1}$
\\
$^{1}$Department of Physics, University of Warwick, Coventry CV4 7AL, UK
}

\begin{document}
\label{firstpage}
\pagerange{\pageref{firstpage}--\pageref{lastpage}}
\maketitle

\begin{abstract}
Although conjunctions and oppositions frequently occur in planetary systems, eclipse-related phenomena are usually described from an Earth-centric perspective. Space missions to different parts of the Solar system, as well as the mounting number of known exo-planets in habitable zones and the possibility of sending featherweight robot spacecraft to them, prompt broader considerations. Here, we derive the geometry of eclipses, transits and occultations from a primarily exo-Earth viewpoint, and apply the formulation to the Solar system and three types of three-body extrasolar planetary systems: with 1 star and 2 planets (Case I), with 2 stars and 1 planet (Case II), and with 1 planet, 1 star and 1 moon (Case III). We derive the general conditions for total, partial and annular eclipses to occur at exo-syzygy, and implement them in each case in concert with stability criteria. We then apply the formalism to the TRAPPIST-1, Kepler-444 and Kepler-77 systems -- the first of which contains multiple potentially habitable planets -- and provide reference tables of both Solar system and TRAPPIST-1 syzygy properties. We conclude by detailing a basic algebraic algorithm which can be used to quickly characterize eclipse properties in any three-body system. 
\end{abstract}


\section{Introduction}

By the time Nicolaus Copernicus postulated that the Earth revolves around the Sun -- and not vice versa -- humans had already been studying eclipses for thousands of years \citep{lizha1998}. Indeed, one of the oldest-known astrophysical phenomena is syzygy, when three or more celestial bodies become co-linear.  Although often defined in terms of the Sun, Moon and Earth only, syzygys could involve other Solar system bodies, and more than three total. \cite{mcdonald1986} computed the frequencies of a variety of Solar system syzygys, and \cite{peagle2016} provided specific examples, such as when the Sun, Venus, Earth, Jupiter and Saturn achieved syzygy in the year 1683, and when the Sun, Venus, Jupiter and Saturn will achieve syzygy in the year 2040.

In the argot of astronomers, the terms {\it eclipse}, {\it transit}, and {\it occultation} have come to represent syzygy in different contexts (e.g. the Sun is eclipsed, whereas exo-planets transit, and stars are occulted). Semantics aside, the geometrical configuration provided by a syzygy yields invaluable information. The Sun's corona is visible from Earth only during particular types of syzygys, the rings around the asteroid Chariklo were detected only because of a different type of syzygy \citep{braetal2014}, and the first putative detection of rings around an exo-moon were made possible because of syzygys \citep{kenmam2015}. In fact, exo-planetary systems containing multiple planets which transit their parent star or stars have the potential to be ``most information-rich planetary systems besides our own solar system'' \citep{raghol2010}. The idea of exo-syzygy is not just a possibility, but a reality: exo-planets like Kepler-1647~b have already achieved syzygy with both of its parent stars and the Earth \citep{kosetal2016}.

Almost ubiquitously, syzygys have been studied by assuming a viewpoint on or close to the Earth's surface. However, the motivation for studying syzygys in a more general context has now received added impetus with (i) the numerous space missions that visit other parts of the Solar system (and beam back data from a variety of viewpoints), and (ii) the febrile desire to find habitable planets and understand what life is like on them. Cassini, MESSENGER and New Horizons are but a few of the robotic spacecraft which have provided us with different perspectives. The great fortune that our nearest stellar neighbour (Proxima Centuri) happens to host a detectable planet in the habitable zone \citep{angetal2016} has provided an ideal flyby candidate for the Breakthrough Starshot mission\footnote{http://breakthroughinitiatives.org/}, which has already received significant financial backing.

Here, we formulate a geometry for three-body syzygys in a general-enough context for wide applications and which yields an easy-to-apply collection of formulae given only the radii and mutual distances of the three bodies. We then apply the formulae to a wide variety of Solar system syzygys and extrasolar syzygys. We begin in Section 2 by establishing our setup. We then place eclipses into context for three different combinations of stars, planets and moons in Sections 3-5 by providing limits in each case from the results of the derivations in Appendices A and B and from stability criteria. We apply our formulae to a plethora of specific cases for the Solar system in Section 6 and for extrasolar systems in Sections 7-8. We conclude with a useful algorithm, including a user-friendly flow chart, in Section 9.

All data used in the applications are taken from two sources unless otherwise specified: the Jet Propulsion Laboratory Solar System Dynamics website\footnote{http://ssd.jpl.nasa.gov/} and the Exoplanet Data Explorer\footnote{http://exoplanets.org/}.

\section{Setup}

We model planetary systems that include three spherical bodies, at least one of which is a star.
Because we consider the systems only at syzygy,
no motion is assumed, allowing for a general treatment with wide applicability.

\subsection{Nomenclature}

We denote the star as the ``primary'', the body on or around which an 
observer/detector might see an eclipse as the ``target'', and the 
occulting/transiting/eclipsing body as the ``occulter''. The occulter
may be a planet (Case I), star (Case II) or moon (Case III), and we 
consider these individual cases in turn in Sections 3-5.  We denote the 
radii of the primary, occulter and target to be, respectively $R_1$, 
$R_2$ and $R_3$, and the pairwise distances between
the centres of the objects as $r_{12}$, $r_{23}$ and $r_{13}$. We assume that
the primary is the largest of the three objects, such that $R_1 > R_2$ and $R_1 > R_3$,
but make no assumptions about the relative sizes of the occulter and target.

All results are derived in terms of the radii and mutual distances only.
The masses of the bodies $M_1$, $M_2$ and $M_3$ rarely factor in the equations, 
but regardless are convenient identifiers for diagrams and will primarily
be used for that purpose.

\subsection{Mutual distances}

The values of any two of $r_{12}$, $r_{13}$ and $r_{23}$ may be given, allowing for
trivial computation of the third value.  For Cases I and II, when the occulter is a planet or star,
the astrocentric distances $r_{12}$ and $r_{13}$ are often known, whereas for Case III,
when the occulter is a moon, $r_{13}$ and $r_{23}$ are often known from observations.

However, in both observational and theoretical studies, orbits with respect to centres
of mass (as in barycentric and Jacobi coordinates) are sometimes more valuable. 
Denote $r_{123}$ as the distance of the target to the centre of mass of the primary and occulter.
Then 

\begin{equation}
r_{13} = r_{123} + \left(\frac{M_1}{M_1 + M_2}\right) r_{12}
. 
\label{r13}
\end{equation}

Observations from Earth rarely catch three-bodies in an exo-planetary system in syzygy.
Therefore, instantaneously measured or estimated mutual distances may
not be as helpful as orbital parameters, such as semimajor axis $a$, eccentricity
$e$ and true anomaly $\Pi$. They are related to separation as

\begin{equation}
r_{12} = \frac{a_{12} \left(1 - e_{12}^2\right)}{1 + e_{12} \cos{\left[\Pi_{12}(t_{\rm s})\right]} }
\label{r12}
\end{equation}

\noindent{}with similar forms for $r_{23}$, $r_{13}$ and $r_{123}$.

The values of $a$ and $e$ can be treated as fixed
on orbital timescales, whereas $\Pi$ is a proxy for time evolution. In three-body
systems, at some points along the mutual orbits -- at nodal intersections --
the three bodies achieve syzygy. The inclinations, longitudes of ascending nodes
and arguments of pericentre
of the bodies help determine eclipse details associated with motion (such as duration length);
here we consider only the static case at syzygy, and hence are unconcerned with 
these other parameters, as well as the time at which the true anomalies achieve syzygy.

\subsection{Radiation cones}

In all cases, the radiation emanating from the primary will form two different types of cones
with the occulter, because the latter is smaller than the former. The first type of cone,
yielding total and annular eclipses,
is formed from outer or external tangent lines (Fig. \ref{cart1}; for derivations, see Appendix A).  The second 
type of cone, yielding partial eclipses, form from the inner or internal tangent 
lines (Fig. \ref{cart2}; for derivations, see Appendix B).

\begin{figure}
\includegraphics[width=8cm]{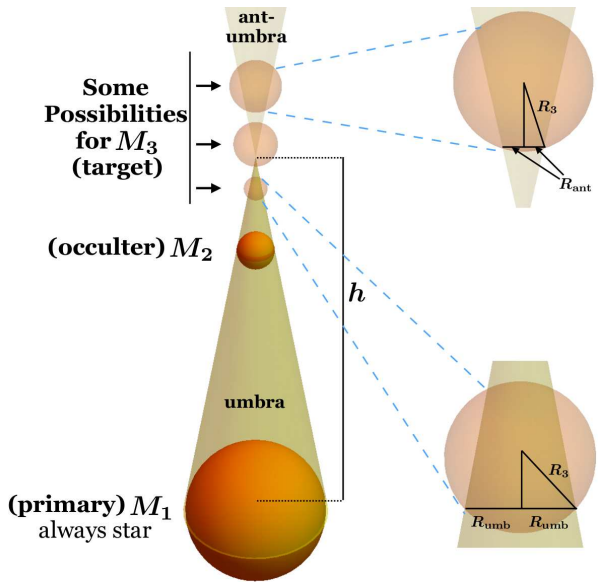}
\caption{
A geometrical sketch for the umbra and antumbra. The cone defined by the radiation 
emitted from the spherical primary
has a height $h$ and circumscribes the occulter. If the target intersects the umbral cone at syzygy,
then a total eclipse will occur as seen by an observer on the target. The projected area of
the total eclipse area is defined by the radius $R_{\rm umb}$. Otherwise, an annular eclipse will
occur, with a projected radius of $R_{\rm ant}$. The primary is always
larger than the occulter and target.
}
\label{cart1}
\end{figure}

\begin{figure}
\includegraphics[width=5cm]{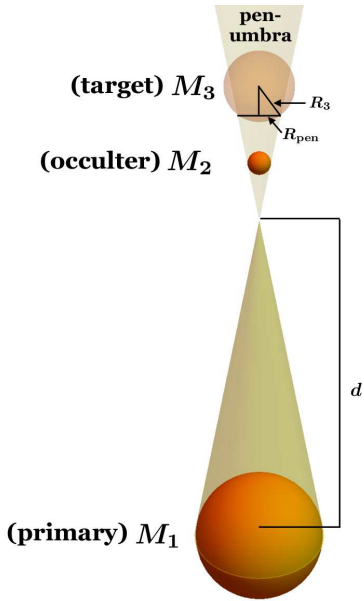}   
\caption{
A geometrical sketch for the penumbral shadow, which,
in contrast to Fig. \ref{cart1}, is bounded by internal
tangent lines between the primary and occulter (the external
tangent lines of Fig. \ref{cart1} produce a longer cone). The height of
the cone here is $d$. The illustration
depicts the shadow falling on only part of the target; for a distant-enough
or small-enough target, $R_{\rm pen}$ would no longer be defined.
} 
\label{cart2}
\end{figure}

\section{Case I: planet occulter}

We now apply our geometrical formalism to three general cases, starting
with that of a planet occulter. One specific example of this case would be Mercury
transiting the Sun as viewed from Earth.  In what follows, we note that
given semimajor axis $a$ and eccentricity $e$, one can 
generate bounds on the mutual distances by considering their pericentric 
$\left[a(1-e)\right]$ and apocentric $\left[a(1+e)\right]$ 
values.

\subsection{Never total eclipses}

From equation (\ref{cond}), a total eclipse can never occur if

\begin{equation}
{\rm max}(h+n) < {\rm min}(r_{13}) - R_3
\label{nevminmax}
\end{equation}

\noindent{}or 

\begin{equation}
\frac{R_1}{R_1 - R_2} < \left(\frac{a_{13}}{a_{12}}\right)
                         \left(\frac{1 - e_{13} - \frac{R_3}{a_{13}}}{1 + e_{12}}\right)
.
\label{rratio}
\end{equation}

In all cases, because the primary is a star and the target
must lie at a sufficiently large distance from the occulter to remain stable,
the term $R_3/a_{13}$ is negligible. For example, in an extreme case, for a 
Jupiter-sized planet at 0.05 au away from a star (with another planet in-between), 
then $R_3/a_{13} \approx 1\%$.

The ratio on the left-hand side of the equation could take on a variety of values
depending on the nature of the star. If the star is a giant branch star, then
$R_1 \gg R_2$ and

\begin{equation}
\left[1 + \frac{R_2}{R_1} \right] a_{12} \left(1 + e_{12}\right) \lesssim a_{13} \left(1 - e_{13} \right)
\label{never}
\end{equation}

\noindent{}which is effectively a condition on crossing orbits and hence is
always true for our setup. {\it Therefore, total eclipses cannot occur in extrasolar 
systems with two planets and one giant star}.

If instead the star is a main sequence star larger than a red dwarf, then it is useful to
consider the fact that the planet's size is bounded according to 
max$(R_2) \approx 0.01R_{\odot}$. For a red dwarf,
around which potentially habitable planets like Proxima b
have been discovered \citep{angetal2016}, max$(R_2) \approx 0.1R_{\odot}$.
In either approximation, {\it for any main sequence stars, equation (\ref{never}) holds and total
eclipses can never occur}. 

\subsection{Always total eclipses}

For smaller stars, like white dwarfs, we now consider the opposite extreme.

A total eclipse will {\it always} occur if

\begin{equation}
{\rm min}(h+n) \ge {\rm max}(r_{13}) - R_3
\label{alwminmax}
\end{equation}

\noindent{}or

\begin{equation}
\frac{R_1}{R_1 - R_2} \ge \left(\frac{a_{13}}{a_{12}}\right)
                         \left(\frac{1 + e_{13} - \frac{R_3}{a_{13}}}{1 - e_{12}}\right)
,
\label{alwear}
\end{equation}

\noindent{}which we approximate as

\begin{equation}
a_{12} \left(1 - e_{12}\right) \gtrsim a_{13} \left(1 + e_{13} \right)
\left[1 - \frac{R_2}{R_1} \right]
.
\label{always}
\end{equation}

\noindent{}by removing the $R_3/a_{13}$ term but otherwise not making any assumptions
about the relative values of $R_1$ and $R_2$. Equation (\ref{always}) holds true when
the term in the square brackets is sufficiently small.

{\it Therefore, if the primary and occulter are about the same size (such as an Earth-like
planet orbiting a white dwarf), then a total eclipse
always occurs}. Otherwise, the bracketed term determines the factor by
which the apocentre of the outer planet must be virtually reduced in 
order to satisfy the equation. Usefully, the equation explicitly contains
the pericentre of the inner planet and the apocentre of the outer planet.

\begin{figure}
\includegraphics[width=8.5cm]{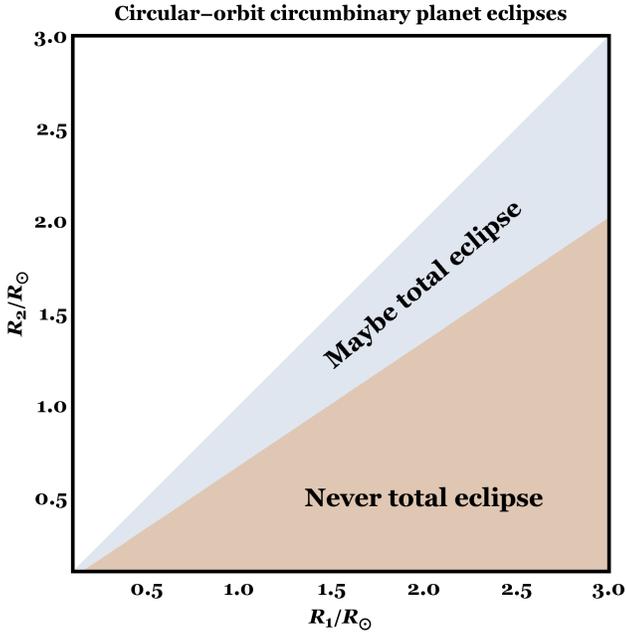}
\caption{
Where total eclipses in circumbinary systems cannot occur.
Here circular orbits ($e_{12} = e_{123} = 0$) and equal 
stellar densities are assumed. The units of $R_1$ and $R_2$
were arbitrarily chosen, and in fact may be any other unit.
This plot was derived from equation (\ref{Rseven}).
}
\label{cont}
\end{figure}

\section{Case II: stellar occulter}

Because circumbinary planets are now known to be common and will continue to represent
a source for some of the most fascinating planetary systems in the foreseeable
future \citep{armetal2014,martri2015,sahetal2015}, the case of a stellar occulter is important to consider.
When a system contains two stars and one circumbinary planet, then the planet's
orbital elements are usually measured with respect to the centre of mass
of the stars (and contain a subscript of ``123'', as indicated in Section 2).

\subsection{Never total eclipses}

Hence, combining the condition for total eclipses to
never occur (equation \ref{nevminmax})
with equation (\ref{r13}) yields

\begin{equation}
{\rm min}(h+n) < {\rm max}(r_{123}) + {\rm max}\left(\frac{r_{12}M_1}{M_1+M_2}\right) - R_3.
\end{equation}

This expression, under the reasonable assumption $e_{12}$~$=$~$0$ for
circumbinary planet host stars, gives

\begin{equation}
  \frac{R_1}{R_1-R_2} < \frac{a_{123}\left(1 - e_{123}\right)}{a_{12}}
  + \frac{M_1}{M_1+M_2} - \frac{R_3}{a_{12}}
  .
  \label{e12circ}
\end{equation}

\noindent{}The last term can be neglected because $R_3 \ll R_1 < a_{12}$.
We can go further and consider cases in which both stars
have the same density. Then equation (\ref{e12circ})
becomes

\begin{equation}
  \frac{R_1R_2 \left(R_{1}^2 + R_{2}^2\right)}
       {\left(R_1 - R_2\right)\left(R_{1}^3 + R_{2}^3\right)}
       <
       \frac{a_{123} \left(1 - e_{123} \right)}{a_{12}},
\end{equation}

\noindent{}which illustrates that to ensure a total eclipse will never
occur, a planet's pericentre must be sufficiently distant, and the stellar
orbit must be sufficiently tight.

We can now take the relation even further. If we know that the planet
is on a circular orbit ($e_{123}=0$),
then we may use the stability boundary of equation 3 of
\cite{holwie1999}, which yields
(still assuming $e_{12} = 0$),

\

\

\[
{\rm min}\left(a_{123}\right) \approx
\]

\begin{equation}
\ \ \ a_{12} \left[
1.60 + 4.12 \left(\frac{M_2}{M_1 + M_2} \right) - 5.09 \left(\frac{M_2}{M_1 + M_2}\right)^2
\right].
\end{equation}

This boundary, along with the assumptions of circularity and equal stellar densities,
allow us to write the condition for ensuring total eclipses on the target in
terms of only the stellar radii as

\[
  \frac{R_{1}^6 R_2 - 3.12 R_{1}^4 R_{2}^3 + 5.12 R_{1}^3 R_{2}^4
  + 1.97R_{1}R_{2}^6 - 0.97R_{2}^7
  }
       {\left(R_1 - R_2\right)\left(R_{1}^3 + R_{2}^3\right)^2}
       < 1.6
       .
\]

\begin{equation}
\label{Rseven}
\end{equation}

\noindent{}We plot Fig. \ref{cont} from equation (\ref{Rseven}), and recall
that our setup requires $R_1>R_2$. The plot emphasizes that the radii of
both stars must be close to each other in order to maintain the possibility
of a total eclipse.

\subsection{Always total eclipses}

In order to guarantee a total eclipse, we combine equation (\ref{alwminmax})
with equation (\ref{r13}) to yield

\begin{equation}
  \frac{R_1}{R_1-R_2} \ge \frac{a_{123}\left(1 + e_{123}\right)}{a_{12}}
  + \frac{M_1}{M_1+M_2} - \frac{R_3}{a_{12}}
  .
  \label{e12tot}
\end{equation}

This equation illustrates that when $R_1 = R_2$, a total eclipse will
always occur. We can again neglect the last term. Under the assumption
of equivalent stellar densities, we find

\begin{equation}
  \frac{R_1R_2 \left(R_{1}^2 + R_{2}^2\right)}
       {\left(R_1 - R_2\right)\left(R_{1}^3 + R_{2}^3\right)}
       \ge
       \frac{a_{123} \left(1 + e_{123} \right)}{a_{12}}.
\end{equation}

\section{Case III: moon occulter}

For a moon occulter (such as the Sun-Earth-Moon system), 
we follow a similar procedure as
to the last two sections in order
to provide relations which establish parameter space regimes
in which total eclipses can never or always occur.

First we note that additional constraints can be imposed
on the moon's orbital size, from both below and above.
These constraints may be used with the formulae below depending
on what quantities are known.
As detailed in \cite{payetal2016}, moons can exist in stable orbits if they
reside somewhere outside of the Roche, or disruption, radius of the planet,
and inside about one-half of a Hill radius (see also 
\citealt*{hamkri1997}, \citealt*{dometal2006} and \citealt*{donnison2010}). Hence,

\begin{equation}
k_{\rho} \left(\frac{M_1}{\rho_2} \right)^{1/3}  < r_{23} < k_{\rm H} a_{13} \left(1 - e_{13}\right) \left(\frac{M_3}{3M_1} \right)^{1/3}
\end{equation}

\noindent{}where $\rho$ refers to density, and $k_{\rho}$ and $k_{\rm H}$ are constants.
$k_{\rho}$ is a constant which can take on a variety of values
depending on the body's shape, composition and spin, and typically span the range 0.78-1.53
\citep[Table 1 of][]{veretal2017}. Although a common value of $k_{\rm H}$ is 0.5, it is
dependent on the direction of revolution and other properties of the moon.

\subsection{Never total eclipses}

In the context of typically-known orbital elements in systems
with a moon, equation (\ref{nevminmax}) becomes

\begin{equation}
  \left(\frac{R_1}{R_1-R_2}\right) {\rm max}\left(r_{13} - r_{23}\right)
  <
  {\rm min}\left(r_{13}\right) - R_3
  ,
\end{equation}

\noindent{}which can be simplified, by neglecting a $R_3/a_{13}$ term,
to

\begin{equation}
  \frac{a_{23}}{a_{13}} > \frac{1+e_{13}-\left(1 - e_{13}\right)
  \left[\frac{R_1 - R_2}{R_1} \right] }
       {1-e_{23}}.
       \label{moon1}
\end{equation}

If both orbits are circular, then equation (\ref{moon1}) reduces
to the compact form

\begin{equation}
  \frac{a_{23}}{a_{13}} > \frac{R_2}{R_1}.
  \label{simple}
\end{equation}

For the case of the Sun, Moon and Earth, both sides of equation (\ref{simple})
are nearly equal (to within a few per cent of about 0.0025). Hence, because the
equation sometimes holds, the Moon sometimes produces total eclipses, and
sometimes does not.

\subsection{Always total eclipses}

In a similar fashion, the condition to ensure that total eclipses will occur
(equation \ref{alwminmax}) gives

\begin{equation}
  \left(\frac{R_1}{R_1-R_2} \right){\rm min}\left(r_{13} - r_{23}\right)
  \ge
  {\rm max}\left(r_{13}\right) - R_3
  ,
\end{equation}

\noindent{}or, by neglecting a $R_3/a_{13}$ term,

\begin{equation}
  \frac{a_{23}}{a_{13}} \le \frac{1-e_{13}-\left(1 + e_{13}\right)
  \left[\frac{R_1 - R_2}{R_1} \right] }
       {1+e_{23}}
       \label{moon1}
       .
\end{equation}

\section{Application to Solar system}

We begin applying our general formalism to real systems by considering
three-body subsets of the Solar system. Table \ref{ectable} lists
some properties of these subsystems. For almost every subsystem,
we provide two extreme cases, when the distance between the occulter
and target are minimized and maximized. We achieve these limits
by computing the relevant ratios of orbital pericentres and apocentres.

We emphasize that the values in the table are theoretical estimates
based on our formalism here and do not take into account the
many complications that exist in reality, such as oblateness
and albedo effects. Nevertheless, the table reveals interesting
facets about eclipses.

\begin{table*}
  \centering
\begin{tabular}{c c c c c c c c c c c}
\label{ectable}
\\
\hline 
\multicolumn{1}{c}{primary}  &
\multicolumn{1}{c}{occulter} &
\multicolumn{1}{c}{target}   &
\multicolumn{1}{c}{$r_{23}$}  &
\multicolumn{1}{c}{$r_{12}$}  &
\multicolumn{1}{c}{umbra or} &
\multicolumn{1}{c}{engulfed} &
\multicolumn{1}{c}{engulfed} &
\multicolumn{1}{c}{$R_{\rm umb}$ or}  &
\multicolumn{1}{c}{max($\phi_{23}$)} &
\multicolumn{1}{c}{max$(g)$}
\\ 
\multicolumn{1}{c}{} &
\multicolumn{1}{c}{} &
\multicolumn{1}{c}{} &
\multicolumn{1}{c}{} &
\multicolumn{1}{c}{} &
\multicolumn{1}{c}{antumbra}    &
\multicolumn{1}{c}{in umb/ant?} &
\multicolumn{1}{c}{in pen?}     &
\multicolumn{1}{c}{$R_{\rm ant}$ (km)} &
\multicolumn{1}{c}{} &
\multicolumn{1}{c}{}
\\
\hline
\\
Sun   &  Mercury  &   Earth  & min   & max      & ant  & yes  &  yes  & -- & 0.217'  &  $4.45 \times 10^{-5}$ \\
Sun   &  Mercury  &   Earth  & max   & min      & ant  & yes  & yes  & -- & 0.158'  &  $2.53 \times 10^{-5}$ \\
Sun   &  Venus  &   Earth  & min   & max      & ant  & yes  & yes  & -- & 1.09'  &  $1.12 \times 10^{-3}$ \\
Sun   &  Venus  &   Earth  & max   & min     & ant  & yes  & yes  & -- & 0.933  &  $8.79 \times 10^{-4}$ \\
Sun   &  Moon  &   Earth  & min   & max      & ant  & no  & no  & 149 & 29.9' & $0.847$ \\
Sun   &  Moon  &   Earth  & max   & min      & umb  & no  & no  & 110 & 31.5'  &  1.0 \\
Jupiter &  Io  &   Earth    &   min   &  max   & ant   &  yes  & yes   & --  & 0.0213'   &  $6.80 \times 10^{-4}$  \\
Jupiter &  Europa  &   Earth    &   min   &  max   & ant   &  yes  & yes   & --  & 0.0183'   &  $5.00 \times 10^{-4}$  \\
Jupiter &  Ganymede  &   Earth    &   min   &  max   & ant   &  yes  & yes   & --  & 0.0308'   &  $1.42 \times 10^{-3}$  \\
Jupiter &  Callisto  &   Earth    &   min   &  max   & ant   &  yes  & yes   & --  & 0.0030'   &  $1.31 \times 10^{-5}$  \\
Sun   &  Earth  &   Moon  & min   & min      & umb  & yes  & yes  & -- & 32.5'  &  1.0 \\
Sun   &  Earth  &   Moon  & max   & max      & umb  & yes  & yes  & -- & 31.4'  &  1.0 \\
Sun   &  Earth  &   Mars  & min   & max      & ant  & yes  & yes  & -- & 0.803'  &  $1.20 \times 10^{-3}$ \\
Sun   &  Earth  &   Mars  & max   & min      & ant  & yes  & yes  & -- & 0.429'  &  $4.99 \times 10^{-4}$ \\
Sun   &  Phobos  &   Mars  & min   & min     & ant  & no  & no  & 9.34 & 12.5'  &  $0.289$ \\
Sun   &  Phobos  &   Mars  & max   & max     & ant  & no  & no  & 5.02 & 13.1'  &  $0.463$ \\
Sun   &  Deimos  &   Mars  & min   & min     & ant  & no  & no  & 61.4 & 2.12'  &  $8.40 \times 10^{-3}$ \\
Sun   &  Deimos  &   Mars  & max   & max     & ant  & no  & no  & 49.8 & 2.12'  &  $1.22 \times 10^{-2}$ \\
Sun   &  Earth  &   Jupiter  & min   & max   & ant  & yes  & yes  & -- & 0.0744'  &  $1.33 \times 10^{-4}$ \\
Sun   &  Earth  &   Jupiter  & max   & min   & ant  & yes  & yes & --  & 0.0655'  &  $1.25 \times 10^{-4}$ \\
Sun   &  Io  &   Jupiter  & min   & min      & umb  & no  & no  & 1490 & 6.46'  &  1.0 \\
Sun   &  Io  &   Jupiter  & max   & max      & umb  & no  & no  & 1520 & 5.86'  &  1.0 \\
Sun   &  Europa  &   Jupiter  & min   & min  & umb  & no  & no  & 990 & 6.46'  &  1.0 \\
Sun   &  Europa  &   Jupiter  & max   & max  & umb  & no  & no  & 1050 & 5.86'  &  1.0 \\
Sun   &  Ganymede  &   Jupiter  & min   & min  & umb  & no  & no  & 1690 & 6.46'  &  1.0 \\
Sun   &  Ganymede  &   Jupiter  & max   & max  & umb  & no  & no  & 1780 & 5.86'  &  1.0 \\
Sun   &  Callisto  &   Jupiter  & min   & min  & umb  & no  & no  & 694 & 6.46'  &  1.0 \\
Sun   &  Callisto  &   Jupiter  & max   & max  & umb  & no  & no  & 877 & 5.86'  &  1.0 \\
Sun   &  Pluto  &   Charon  & circ   & min   & umb  & yes  & yes  & -- & 1.08'  &  1.0 \\
Sun   &  Pluto  &   Charon  & circ  & max   & umb  & yes  & yes  & -- & 0.649'  &  1.0 \\
Sun   &  Charon  &   Pluto  & circ   & min  & umb  & no  & no  & 603 & 1.08'  &  1.0 \\
Sun   &  Charon  &   Pluto  & circ   & max  & umb  & no  & no  & 604 & 0.649'  &  1.0 \\
Sun   &  Jupiter  &   Pl. Nine  & circ   & min   & ant  & yes  & yes  & -- & 0.0043'  &  0.010 \\
Jupiter &  Io  &   Europa    &   min   &  max   & ant   &  yes  & yes   & --  & 52.3'   &  $5.18 \times 10^{-3}$  \\
Jupiter &  Io  &   Europa    &   max   &  min   & ant   &  yes  &  yes  & --  & 48.9'   &  $4.73 \times 10^{-3}$  \\
Jupiter &  Io  &   Ganymede    &  min    &  max   & ant   & yes   &  yes  & --  & 19.5'   &  $1.87 \times 10^{-3}$  \\
Jupiter &  Io  &   Ganymede    &  max    &  min   & ant  &  yes &  yes  & --  &  19.3'  &  $1.84 \times 10^{-3}$  \\
Jupiter &  Ganymede  &  Callisto    &  min    & max    & ant   & yes   & yes   & --  & 22.8'  & $7.81 \times 10^{-3}$   \\
Jupiter &  Ganymede  &  Callisto    &  max    & min    & ant   & yes   & yes   & --  &  21.9'  &  $7.46 \times 10^{-3}$  \\
Saturn &  Encleadus  &  Tethys    &  circ   &  circ   & ant   & yes   &  yes  & --  &  30.9'  &  $5.09 \times 10^{-4}$  \\
Saturn &  Titan  &  Hyperion    &   min   &  max   &  ant  &  yes  &  yes  & --  &  84.7'  &  $9.62 \times 10^{-2}$  \\
Saturn &  Titan  &  Hyperion    &   max   &  min   &  ant  &  yes  &  yes  & --  &  50.7'  &  $3.78 \times 10^{-2}$  \\
Uranus &  Titania  &  Oberon    &   min   &  max   &  ant  &  yes  &  yes  & --  &  37.4'  &  $1.55 \times 10^{-2}$  \\
Uranus &  Titania  &  Oberon    &   max   &  min   &  ant  &  yes  &  yes  & --  &  36.7'  &  $1.51 \times 10^{-2}$  \\
Neptune &  Triton  &  Nereid    &   circ   &  max   &  ant  &  yes  &  yes  & --  &  9.12'  &  $5.49 \times 10^{-3}$  \\
Neptune &  Triton  &  Nereid    &   circ   &  min   &  ant  &  yes  &  yes  & --  & 1.00'  &  $3.25 \times 10^{-3}$  \\
\end{tabular}
\caption{Eclipse properties of three-body subsystems within the Solar system at syzygy. A value of ``circ'' in the fourth and fifth columns indicates a circular orbit.  The penultimate column provides the maximum possible angular diameter as observed on the target, and the final column gives the maximum eclipse depth (from equation \ref{gval}). When this latter value is equal to unity, a total eclipse occurs for an observer co-linear with the syzygy.}
\end{table*}

\begin{figure}
\includegraphics[width=8.5cm]{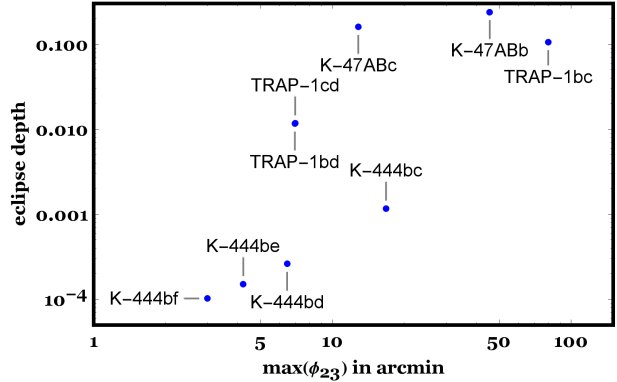}
\caption{
The eclipse depths and angular diameters of some exo-planetary syzygys in
the TRAPPIST-1, Kepler-444 and Kepler-47 systems. Kepler-47 is a circumbinary
system and the other two are single-star systems. The letters indicate
which bodies are in syzygy.
}
\label{Exostats}
\end{figure}

\begin{table*}
  \centering
\begin{tabular}{c c c c c c c c c c c}
\label{ectable}
\\
\hline 
\multicolumn{1}{c}{primary}  &
\multicolumn{1}{c}{occulter} &
\multicolumn{1}{c}{target}   &
\multicolumn{1}{c}{$r_{23}$}  &
\multicolumn{1}{c}{$r_{12}$}  &
\multicolumn{1}{c}{umbra or} &
\multicolumn{1}{c}{engulfed} &
\multicolumn{1}{c}{engulfed} &
\multicolumn{1}{c}{$R_{\rm umb}$ or}  &
\multicolumn{1}{c}{max($\phi_{23}$)} &
\multicolumn{1}{c}{max$(g)$}
\\ 
\multicolumn{1}{c}{} &
\multicolumn{1}{c}{} &
\multicolumn{1}{c}{} &
\multicolumn{1}{c}{} &
\multicolumn{1}{c}{} &
\multicolumn{1}{c}{antumbra}    &
\multicolumn{1}{c}{in umb/ant?} &
\multicolumn{1}{c}{in pen?}     &
\multicolumn{1}{c}{$R_{\rm ant}$ (km)} &
\multicolumn{1}{c}{} &
\multicolumn{1}{c}{}
\\
\hline
\\
TRAPPIST-1  &  Planet b  &   Planet e  & circ   & circ      & ant  & yes  &  yes  & -- & 18.7'  &  $1.96 \times 10^{-2}$ \\
TRAPPIST-1  &  Planet c  &   Planet e  & circ   & circ      & ant  & yes  &  yes  & -- & 23.9'  &  $3.18 \times 10^{-2}$ \\
TRAPPIST-1  &  Planet d  &   Planet e  & circ   & circ     & ant  & yes  &  yes  & -- & 32.5'  &  $5.89 \times 10^{-2}$ \\
TRAPPIST-1  &  Planet b  &   Planet f  & circ   & circ     & ant  & yes  &  yes  & -- & 12.3'  &  $1.46 \times 10^{-2}$ \\
TRAPPIST-1  &  Planet c  &   Planet f  & circ   & circ      & ant  & yes  &  yes  & -- & 14.1'  &  $1.93 \times 10^{-2}$ \\
TRAPPIST-1  &  Planet d  &   Planet f  & circ   & circ      & ant  & yes  &  yes  & -- & 14.2'  &  $1.96 \times 10^{-2}$ \\
TRAPPIST-1  &  Planet e  &   Planet f  & circ   & circ      & ant  & yes  &  yes  & -- & 30.0'  &  $8.78 \times 10^{-2}$ \\
TRAPPIST-1  &  Planet b  &   Planet g  & circ   & circ      & ant  & yes  &  yes  & -- & 9.37'  &  $1.27 \times 10^{-2}$ \\
TRAPPIST-1  &  Planet c  &   Planet g  & circ   & circ      & ant  & yes  &  yes  & -- & 10.3'  &  $1.54 \times 10^{-2}$ \\
TRAPPIST-1  &  Planet d  &   Planet g  & circ   & circ      & ant  & yes  &  yes  & -- & 9.44'  &  $1.28 \times 10^{-2}$ \\
TRAPPIST-1  &  Planet e  &   Planet g  & circ   & circ      & ant  & yes  &  yes  & -- & 15.9'  &  $3.63 \times 10^{-2}$ \\
TRAPPIST-1  &  Planet f  &   Planet g  & circ   & circ      & ant  & yes  &  yes  & -- & 38.5'  &  $2.14 \times 10^{-2}$ \\
\end{tabular}
\caption{Eclipse properties of three potentially habitable planets (planets e, f and g) in the TRAPPIST-1 system. The columns are equivalent to those in Table 1 for easy comparison; the penultimate column provides the maximum possible angular diameter as observed on the target, and the final column gives the maximum eclipse depth (from equation \ref{gval}). All planets are assumed to be coplanar and have circular orbits. Data for these computations is taken from Gillon et al. (2017).}
\end{table*}

\subsection{Observers on Earth}

The first ten cases are applicable to observers on Earth.
The disc of Venus appears 5-10 times larger than the disc
of Mercury when transiting the Sun. Mercury's angular diameter
can change significantly -- by about a quarter -- depending on
its distance from both the Sun and the Earth.
Both Mercury and Venus 
generate transit depths which are less than $10^{-3}$,
and antumbral cones which completely engulf the Earth.

The Sun-Earth-Moon cases emphasize how unusual the
situation is in which humans find themselves. Depending on
the Sun-Earth and Earth-Moon distances, either
a total eclipse or annular eclipse can occur\footnote{In rare
cases, a ``hybrid'' or ``mixed'' eclipse occurs when
both total and annular eclipses occur during the same event.
More technically, the transition into and out of syzygy creates changes
in all mutual distances such that during this transition
some parts of the Earth satisfy the upper branch
of equation (\ref{cond}) and others satisfy the lower branch of that
equation.}.
The minimum eclipse depth is about 0.85, which demonstrates
that at a minimum, 92 per cent of the Sun's disc is covered
by the Moon. Neither the umbral nor antumbral cones ever
engulf the Earth. In fact, the projected radius of the shadow 
cast on Earth does not exceed 150 km (or 93 
miles)\footnote{Actually a more accurate upper limit, due
to effects not considered here, is 140 km.}.
The penumbral cone also does not engulf the Earth, and
produces a much larger shadow with a projected radius of
3380-3640 km (2100-2260 miles).
Compared to every other table entry with the Sun as primary, 
the angular diameter of the moon on the
Sun's disc is large: about half of a degree. Only observers
on Mars watching Phobos transit the Sun would see an angular
diameter which is a third of that of the Moon on the Sun.
If an astronaut was standing on the Moon during an Earth
eclipse of the Sun, they would always see a total eclipse.

\subsection{Other perspectives}

Observatories on Mars must fit themselves into smaller targets
-- on the order of just km or tens of km -- in order to see annular
eclipses produced by that planet's satellites. 
Phobos' eclipses are substantial enough to block out
about one-third to one-half of the Sun's light.
The eclipses due to the Earth as viewed on Mars are less
prominent, and are roughly comparable to Venus transits
of the Sun as viewed on Earth.

Regarding the outer Solar system,
Jupiter is large enough to easily create 
total eclipses on all of the Galilean satellites. 
Pluto and Charon
create total eclipses on one another, with the umbral 
shadow of Pluto completely engulfing Charon. If Planet Nine
\citep{deldel2014,iorio2014,batbro2016} 
exists (assuming a radius of $1.5R_{\oplus}$ and a separation
of 750 au), then the
view from that distant
wanderer during a Jupiter eclipse would see a transit
depth of 1 per cent, despite an angular diameter of
just one-quarter of an arcsecond.

The last 13 rows of the table describe situations where
Jupiter, Saturn, Uranus or Neptune is considered
to be the primary and the occulter and target are two
attached moons. In all cases, an annular eclipse occurs,
and the antumbral shadow encompasses the target.
The transit depths are within an order of $10^{-3}$
and the angular diameters are on the order of tens of
arcminutes. The largest value of $\phi_{23}$ (over one degree) 
occurs with Titan and Hyperion, which are highly disparate
in size (Titan is 19 times greater in radius) and yet very 
close to each other (in fact within a $4$:$3$ mean motion resonance).

\section{Application to known exo-systems}

Exo-planetary systems are now known to host a wide variety of exo-planets, with masses
ranging from 1.6 Lunar masses \citep{wolszczan1994,konwol2003} all the way into the stellar regime.
About 41 per cent of all known planets are in multiple-planet systems, which lead to eclipses
of Type I. A couple dozen planets have two stellar hosts and orbit in a circumbinary fashion,
leading to Type II eclipses. Although no exo-moons have yet been confirmed (for Type III eclipses), 
vigorous searches have been undertaken \citep{heller2017}, 
even despite our current unfortunate observational sensitivity to a distance range of tenths of au
(where we do not see any moons in the Solar system).

The data we have on exo-planets is typically not as accurate as those for the Solar system planets,
and contains missing parameters. For example, exo-planets detected by transit photometry but not confirmed
with Doppler radial velocity measurements do not have measured masses. Exo-planets detected by radial velocity
measurements but not with transit have no radius measurements, and only lower bounds on mass. Transiting
exo-planets sometimes do not have secure eccentricity values, meaning that their apocentric and pericentric
distances are uncertain. Further, other physical effects which could have significant impacts
on transits (but not treated here) such as oblateness, tidal distortions,
and extended atmospheres are often or always not well-constrained in exoplanetary systems.

Despite these caveats, we can analyze some systems of interest: TRAPPIST-1 \citep{giletal2017}, 
Kepler-444 \citep{cametal2015} and Kepler-47 \citep{oroetal2012}.  TRAPPIST-1 and Kepler-444
are single-star 7-planet and 5-planet systems, containing all terrestrial/rocky planets of radii 
between about $0.4-1.2R_{\oplus}$. Kepler-47 is a 2-planet circumbinary planetary system
with two planets of radii $3.0R_{\oplus}$ and $4.7R_{\oplus}$. Conveniently, all planets in all
systems are lettered in alphabetical order according to their distance from the parent star(s).

By using semimajor axis values to represent distances, we compute some quantities of interest.
Figure \ref{Exostats} displays the maximum eclipse depth ($g$) versus the maximum angular diameter ($\phi_{23}$)
for three-body combinations in these systems. The two highest eclipse depths both arise from
the circumbinary Kepler-47 system, and viewers on the outermost planet (d) in the TRAPPIST-1 system 
will see very similar transit features from each of the inner planets (b and c). In that system, the transit
of planet b as viewed from planet c has the largest angular diameter: over one degree.

We take a closer look at the TRAPPIST-1 system in Table 2, where we compute eclipse quantities
for the three targets which are most likely to be habitable (planets e, f and g). In no case on these
planets will a total eclipse occur, and in all cases the eclipse depth is between one and ten per cent
and the maximum angular diameter between about ten and thirty arcseconds.

\begin{figure*}
\includegraphics[width=15cm]{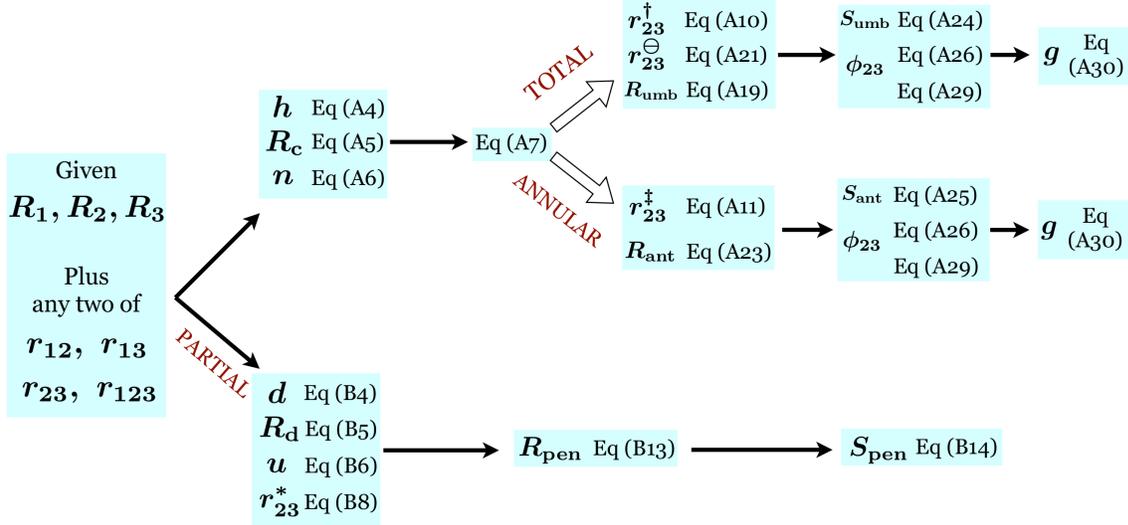}
\caption{
Algorithm to compute eclipse properties given only radii and mutual distances.
}
\label{Flowchart}
\end{figure*}

\section{Application to anticipated exo-systems}

The discovery of exo-moons will inevitably introduce questions about whether and
what types of eclipses will occur. The formalism here can be applied directly to
such situations. As a proof-of-concept, imagine that a large moon was found orbiting
a Jupiter in a low-mass star system, such that $R_1 = 0.2R_{\odot}$, 
$R_2 = R_{\rm Jup}$ and $R_3 = 0.5R_{\oplus}$, with $r_{12} = 1$~au
and $r_{23} = 385 \times 10^3$~km (the mean Moon-Earth distance). In this case, 
the moon would experience total
eclipses, and, as viewed from the moon, max$\left(\phi_{23}\right) \approx 6.4'$.

Another type of planetary system include planets orbiting stars at the point
during their evolution -- the asymptotic giant branch phase -- when the host star 
radius is maximum.  So far, only one planet candidate has been reported to
orbit such a star \citep{keretal2016}, but only after the stellar radius has
contracted following a period of expansion. Suppose instead that the star had $R_1 = 1.5$~au
and hosted two Earth-like planets ($R_2 = R_{\oplus}$ and $R_3 = R_{\oplus}$).
Suppose these planets were sufficiently far ($r_{12} = 2.5$~au and $r_{23} = 3.0$~au) 
to have survived the tidal pull of the star on the asymptotic giant branch \citep{musvil2012}. Then 
an annular eclipse would occur, and despite the relatively enormous stellar 
size, max$(\phi_{23}) = 0.59'$. However,
$g = 2.6 \times 10^{-8}$, producing a negligible change in brightness of the giant star.

Now consider white dwarf planetary systems. At least 5 of the Solar system planets will survive
into the Solar white dwarf phase \citep{veras2016a,veras2016b}, and signatures of planetary debris persist
in the atmospheres of over 30 per cent of all white dwarfs \citep{koeetal2014} and is detected in
circumstellar discs \citep{farihi2016}. Planets are hence expected to orbit white dwarfs at distances
of a few au. A white dwarf hosting two Earth-like planets then feature syzygys where all three objects
are nearly the same size. Imagine that $R_1 = R_{\oplus}$, $R_2 = 0.95 R_{\oplus}$, $R_3 = 0.95 R_{\oplus}$,
$r_{12} = 2.5$~au and $r_{13} = 3.0$~au. Then a total eclipse will occur such that even though 
the umbral shadow does not quite engulf the target ($R_{\rm umb} = 5990$ km), the penumbral shadow does.
Despite the total eclipse, the maximum angular diameter which would be eclipsed is just 0.97'.

\section{Summary}

Syzygys are common features of planetary systems, including our own. Here,
we have constructed an infundibuliform geometry which characterises eclipses, transits
and occultations under one framework. Given only radii and mutual distances, several
quantities -- such as eclipse type, shadow radii and transit depth -- can be computed with basic
trigonometry (Appendices A and B). We illustrate a user-friendly equation flow detailing these computations
in Fig. \ref{Flowchart}.

We have also analyzed three classes of special cases, when the occulter is
a planet (Section 3), star (Section 4) and moon (Section 5). We found that
total eclipses cannot occur in two-planet systems with a main sequence
or giant branch parent star, but may occur often for two-planet systems containing
a white dwarf. In circumbinary systems with circular orbits and similarly dense stars,
by appealing to stability criteria we have parametrized the condition for total
eclipses to never occur to just two variables: the stellar radii (equation \ref{Rseven}). 
The relations in this paper represent useful estimates which may be employed in mission planning
and for added perspective on a wide variety of exo-systems.

\section*{Acknowledgements}

We thank the referee for helpful comments on the manuscript. DV has received funding from the European Research Council under the European Union's Seventh Framework Programme (FP/2007-2013)/ERC Grant Agreement n. 320964 (WDTracer).

\appendix

\

\

\

\noindent{}In Appendices A and B, we present the geometries which form the basis of the            
results of this paper, and derive key equations used in the main body of the text.

\section{Total and annular eclipses} \label{totanu}

Figure \ref{cart1} displays a schematic for the cone formed from external tangent lines.
If the target at some point along its orbit intersects this cone, then the target is 
in the umbra and a total eclipse will occur. 
Otherwise, the target will be in the antumbra and an annular eclipse will occur.  

\begin{figure}
\ \ \ \ \ \ \ \ \ \ \ \ \ \ \ \ \ \ \ \ \
\includegraphics[width=3.5cm]{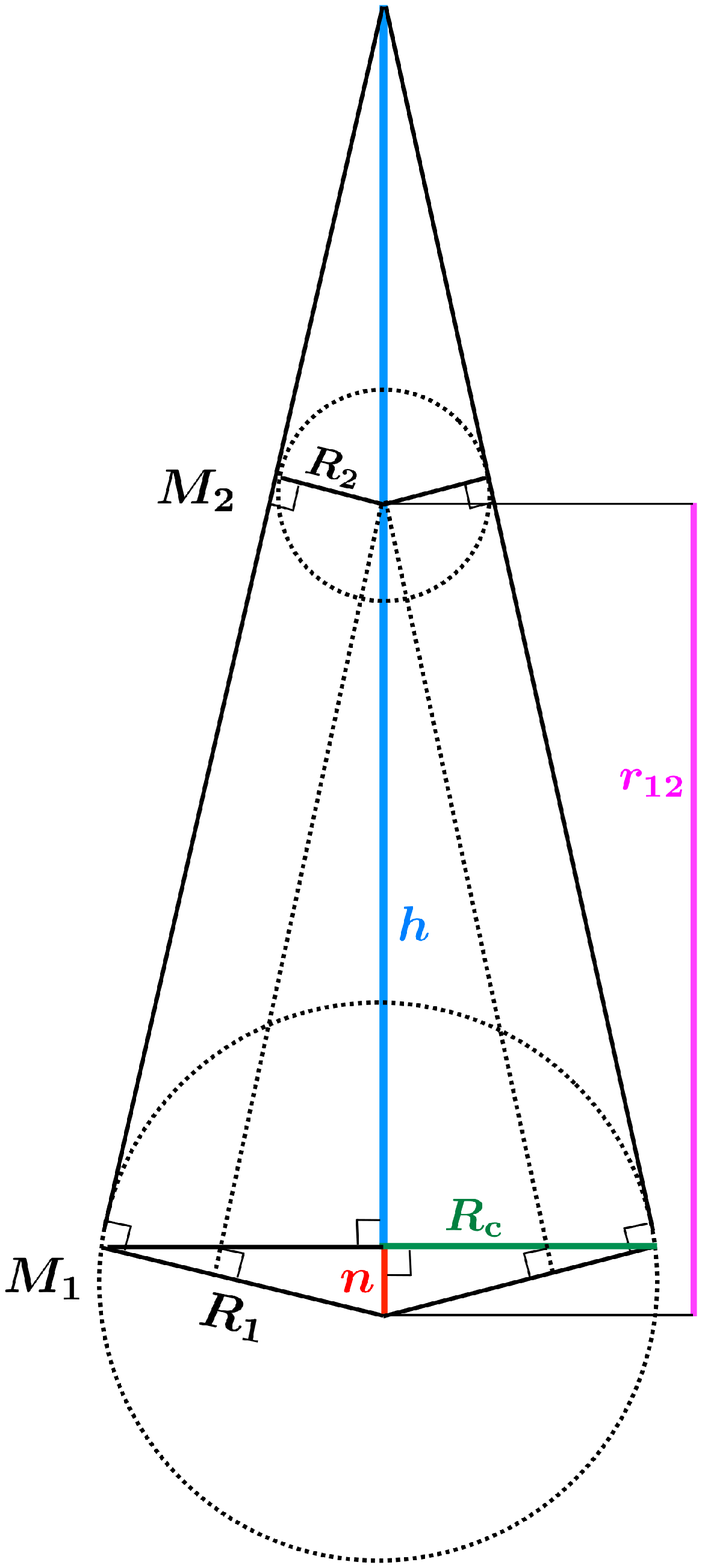}
\caption{
Zoom-in of the umbral cone from Fig. \ref{cart1}. The radius of the
base of the cone is $R_{\rm c}$, and its height is $h$. Equations
(\ref{handn}-\ref{Rcone}) were derived based on the geometry in this diagram.
}
\label{zoomcart}
\end{figure}

Our first goal is to find the height, $h$, of the cone in terms of the given variables
(the radii $R$ of all three objects and their centre-to-centre distances $r$). 
To do so, first consider Fig. \ref{zoomcart}. The location
of the base of the cone is offset from the centre of the primary by a distance
$n$, and the radius of the cone base is $R_{\rm c}$.  Trigonometry gives

\begin{eqnarray}
n     &=& \frac{R_{\rm c}^2}{h}
\label{handn}
\\
h + n &=& \frac{R_1r_{12}}{R_1 - R_2},
\\
R_{\rm c}^2 &=& \frac{1}{2} \left[h\sqrt{h^2 + 4 R_{1}^2} - h^2\right]
\label{Rcone}
\end{eqnarray}

\noindent{}such that their simultaneous solution yields a cone height of

\begin{equation}
h = R_1 \left[\frac{r_{12}}{R_1 - R_2} - \frac{R_1 - R_2}{r_{12}} \right]
.
\label{coneh}
\end{equation}

\noindent{}Consequently, the base of the cone is

\begin{equation}
R_{\rm c} = \left(\frac{R_1}{r_{12}}\right)
\sqrt{
r_{12}^2 - \left(R_1 - R_2 \right)^2
}
\end{equation}

\noindent{}and

\begin{equation}
n = \frac{R_1}{r_{12}} \left(R_1 - R_2\right)
.
\label{conen}
\end{equation}

\noindent{}Note that in equations (\ref{coneh}-\ref{conen}), as $R_2 \rightarrow R_1$, we obtain 
the expected results $n \rightarrow 0$, $R_{\rm c} \rightarrow R_1$ and $h \rightarrow \infty$. 

\subsection{Eclipse criteria}

Either a total or annular eclipse will occur at syzygy depending on the
location of the target with respect to the umbral cone. If the target is beyond
the tip of the cone, then the eclipse is annular. Consequently, 
the condition may be expressed in variables as

\

\[
h  < r_{13} - n - R_3, \ {\rm annular \ eclipse}
\]
\begin{equation}
h \ge r_{13} - n - R_3, \ {\rm total \ eclipse}
\label{cond}
\end{equation}


\begin{figure}
\underline{\ \ \ \ \ \ \ \ \ \ \ UMBRAL LIMIT \ \ \ \ \ \ \ \ \ \ \  ANTUMBRAL LIMIT} 
\includegraphics[width=8cm]{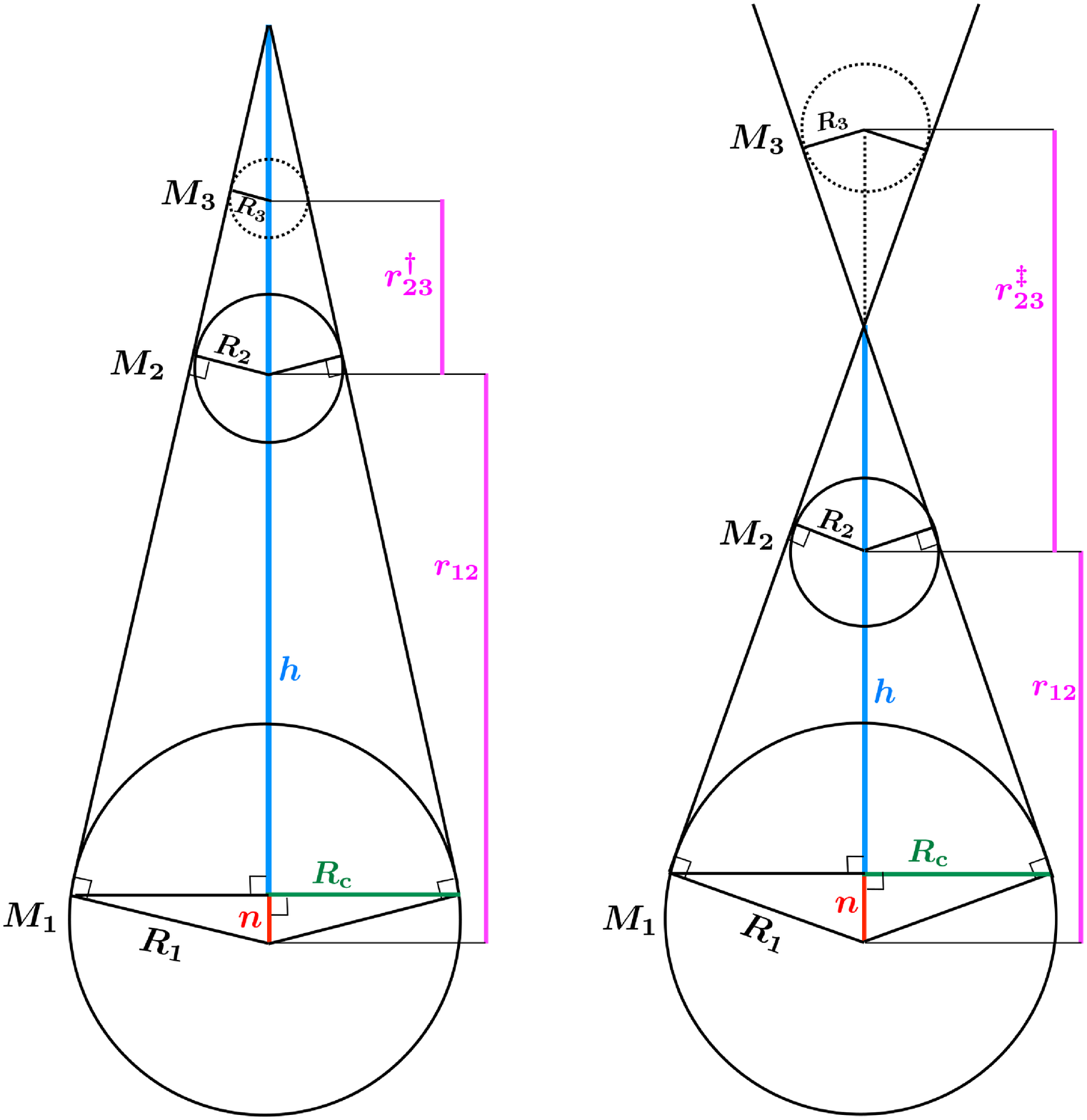}
\caption{
The limiting target-occulter distances ($r_{23}^{\dagger}$ and $r_{23}^{\ddagger}$) 
for which the entire target is covered in the umbra (left-hand panel)
or antumbra (right-hand panel). For $r_{23} \le r_{23}^{\dagger}$, the target
is entirely engulfed in the umbra. For $r_{23} \ge r_{23}^{\ddagger}$, the target
is entirely engulfed in the antumbra.  Equations (\ref{simtri1}-\ref{r3dag})
were derived from the geometry in this figure.
}
\label{limits1}
\end{figure}

\subsection{Shadow radii}

\subsubsection{Total coverage}

In order to determine if the umbral or antumbral shadows will envelope the target
completely, consider Fig. \ref{limits1}, which presents limiting cases for a
static occulter given that the value of $r_{12}$ is fixed.
For the umbral case, let $r_{23}^{\dagger}$ represent the critical value of
$r_{23}$ within which ($r_{23} \le r_{23}^{\dagger}$) the target is engulfed 
in the total eclipse. For
the antumbral case, denote
$r_{23}^{\ddagger}$ as the value beyond which ($r_{23} \ge r_{23}^{\ddagger}$) the 
entire target is engulfed in the annular eclipse. By
entirely engulfed, we are referring to when the entire 
target is within the umbral or antumbral cone.

Similar triangles from the left-hand panel give

\begin{equation}
\frac{R_1}{h+n} 
= \frac{R_2}{h+n-r_{12}}
= \frac{R_3}{h+n-r_{12}-r_{23}^{\dagger}}
\label{simtri1}
\end{equation}

\noindent{}and

\begin{equation}
\frac{R_1}{h+n} 
= \frac{R_2}{h+n-r_{12}}
= \frac{R_3}{-\left(h+n-r_{12}-r_{23}^{\ddagger}\right)}
,
\end{equation}

\noindent{}yielding

\begin{equation}
r_{23}^{\dagger} = r_{12} \left( \frac{R_2 - R_3}{R_1 - R_2} \right)
,
\label{dag1}
\end{equation}

\begin{equation}
r_{23}^{\ddagger} = r_{12} \left( \frac{R_2 + R_3}{R_1 - R_2} \right)
.
\label{dag2}
\end{equation}

Equation (\ref{dag1}) demonstrates that in order for a total eclipse to engulf
the entire target, it must be smaller than the occulter. Alternatively,
Equation (\ref{dag2}) illustrates that any target can be entirely engulfed
by the antumbral shadow if sufficiently far from the occulter. 

These equations may be turned around to solve for any of the other variables
given all others. For example, for a given value of $r_{23}$, the critical
target radii for umbral and antumbral engulfment are

\begin{equation}
R_{3}^{\dagger} = R_2 - \frac{r_{23}}{r_{12}} \left(R_1 - R_2\right),
\label{rdag}
\end{equation}

\begin{equation}
R_{3}^{\ddagger} = \frac{r_{23}}{r_{12}} \left(R_1 - R_2\right) - R_2
.
\label{r3dag}
\end{equation}

At $r_{23}^{\dagger}$ or $R_{3}^{\dagger}$, {\it more} than half of the surface
of the target will be exposed to the umbral shadow. The projected shadow radius $R_{\rm umb}$ 
(see Fig. \ref{cart1}) in this specific case (Fig. \ref{limits1}) is given by

\begin{equation}
\cos{\left(\sin^{-1}\left[\frac{R_{\rm umb}^{\dagger}}{R_3}\right]\right)} = \frac{R_1}{h+n}
\end{equation}

\noindent{}or

\begin{equation}
R_{\rm umb}^{\dagger} = \left(\frac{R_3}{r_{12}}\right)
\sqrt{
r_{12}^2 - \left(R_1 - R_2 \right)^2
}
=
\frac{R_{\rm c} R_3}{R_1}
\end{equation}

In contrast, at $r_{23}^{\ddagger}$ or $R_{3}^{\ddagger}$, {\it less} than half of the surface of the 
target will be exposed to the antumbral shadow. However, geometry illustrates that the 
projected radius of the shadow $R_{\rm ant}$ takes the same form as $R_{\rm umb}$ and is

\begin{equation}
  R_{\rm ant}^{\ddagger} = R_{\rm umb}^{\dagger}
  .
\end{equation}

\subsubsection{Some umbral coverage}

Now we derive more general formulae for $R_{\rm umb}$ and $R_{\rm ant}$ which are not
bound to the geometry in Fig. \ref{limits1}. We begin with the umbral case.  
Consider when $h \ge r_{13} - n - R_3$ and $r_{23} > r_{23}^{\dagger}$, 
such that a total eclipse occurs and the target is not completely engulfed 
in the umbra, as in Fig. \ref{cart1}. If the base of the cone is flush with 
the $x$-$y$ Cartesian plane, then the equation of a double cone (extending
in opposite directions) would be

\begin{equation}
\left(\frac{h}{R_{\rm c}}\right)^2 \left(x^2 + y^2\right) = \left(z - h\right)^2
\label{coneeq}
\end{equation}

\noindent{}so that if a total eclipse occurred on the target, the radius of the
projected area of totality, $R_{\rm umb} = \sqrt{x^2 + y^2}$, would be 

\begin{equation}
R_{\rm umb} = R_{\rm c} \left(1 - \frac{r_{13} - n - \sqrt{R_{3}^2 - R_{\rm umb}^2}}{h}\right)
.
\label{rumbgen}
\end{equation}

\noindent{}Solving for $R_{\rm umb}$ gives

\[
R_{\rm umb} = \left( \frac{h R_{\rm c}}{h^2 + R_{\rm c}^2} \right)
             \bigg[
                    h+n-r_{13}
\]

\begin{equation}
\ \ \ \ \ \ \ \ \ \ \
+
\sqrt{R_{3}^2
-
\left(\frac{R_{\rm c}}{h}\right)^2
\left[
\left( h+n-r_{13} \right)^2 - R_{3}^2
\right]
}
\Bigg]
\end{equation}

\noindent{}which is consistent with the condition given by equation (\ref{cond}).

Also, recall that when the target is completely engulfed in the umbral shadow, then 
over half of the surface is exposed to this shadow, such that $R_3 > R_{\rm umb}^{\dagger}$.  
What if the target is just large enough so that $R_3 = R_{\rm umb}^{\dagger}$? Define
$R_{\rm umb}^{\ominus}$ as the value of $R_3$ such that $R_3 = R_{\rm umb}^{\dagger}$,
and $r_{23}^{\ominus}$ as the value of $r_{23}$ such that $R_3 = R_{\rm umb}^{\dagger}$.
Then equation (\ref{rumbgen}) reduces to

\begin{equation}
R_{\rm umb}^{\ominus} = \left( \frac{R_{\rm c}}{h} \right) \left( h + n - r_{13} \right)
\end{equation}

\noindent{}or, alternatively,

\begin{equation}
r_{23}^{\ominus} = \frac{h}{R_{\rm c}} \left(R_{\rm c} - R_3 \right) + n - r_{12}
.
\end{equation}

\noindent{}The value of $r_{23}^{\ominus}$ represents a useful quantity to access when computing the surface area covered by the shadow.

\subsubsection{Some antumbral coverage}

Now we find a general formula for $R_{\rm ant}$. If an annular eclipse occurs and 
$r_{23} < r_{23}^{\ddagger}$, then the equation of the appropriate cone in this instance yields

\begin{equation}
R_{\rm ant} = R_{\rm c} \left(\frac{r_{13} - n - \sqrt{R_{3}^2 - R_{\rm ant}^2}}{h} - 1\right)
\end{equation}

\noindent{}or

\[
R_{\rm ant} = \left( \frac{h R_{\rm c}}{h^2 + R_{\rm c}^2} \right)
             \bigg[
                    r_{13}-h-n
\]

\begin{equation}
\ \ \ \ \ \ \ \ \ \ \
-
\sqrt{R_{3}^2
-
\left(\frac{R_{\rm c}}{h}\right)^2
\left[
\left( r_{13}-h-n \right)^2 - R_{3}^2
\right]
}
\Bigg]
\end{equation}

\noindent{}which is also consistent with the condition given by equation (\ref{cond}).

\subsection{Shadow surface areas}

Due to geometry of spherical caps, the surface area of the target which is eclipsed 
at syzygy in the umbra, $S_{\rm umb}$ is

\[
S_{\rm umb} = \pi \left[R_{\rm umb}^2 + \left(R_{3} - \sqrt{R_{3}^2 - R_{\rm umb}^2}\right)^2 \right],
\]

\[
\ \ \ \ \ \ \ \ \ \ \ \ \ r_{23} \ge r_{23}^{\ominus};
\]

\[
\ \ \ \ \ \ \ \, = 4\pi R_{3}^2 - \pi \left[R_{\rm umb}^2 + \left(R_{3} - \sqrt{R_{3}^2 - R_{\rm umb}^2}\right)^2 \right],
\]

\begin{equation}
\ \ \ \ \ \ \ \ \ \ \ \ \ r_{23}^{\dagger} \le r_{23} < r_{23}^{\ominus}.
\end{equation}

\noindent{}For $r_{23} < r_{23}^{\dagger}$, the entire target is engulfed and is no longer defined. In the antumbra,

\begin{equation}
S_{\rm ant} = \pi \left[R_{\rm ant}^2 + \left(R_{3} - \sqrt{R_{3}^2 - R_{\rm ant}^2}\right)^2 \right],
\ \ \    r_{23} \le r_{23}^{\ddagger}
\end{equation}

\noindent{}such that for $r_{23} > r_{23}^{\ddagger}$, $R_{\rm ant}$ is no longer defined.

\begin{figure}
\underline{\ \ CO-LINEAR OBSERVER \ \ \ \ \ \ \ \ \ \ \  OFFSET OBSERVER} 
\includegraphics[width=8cm]{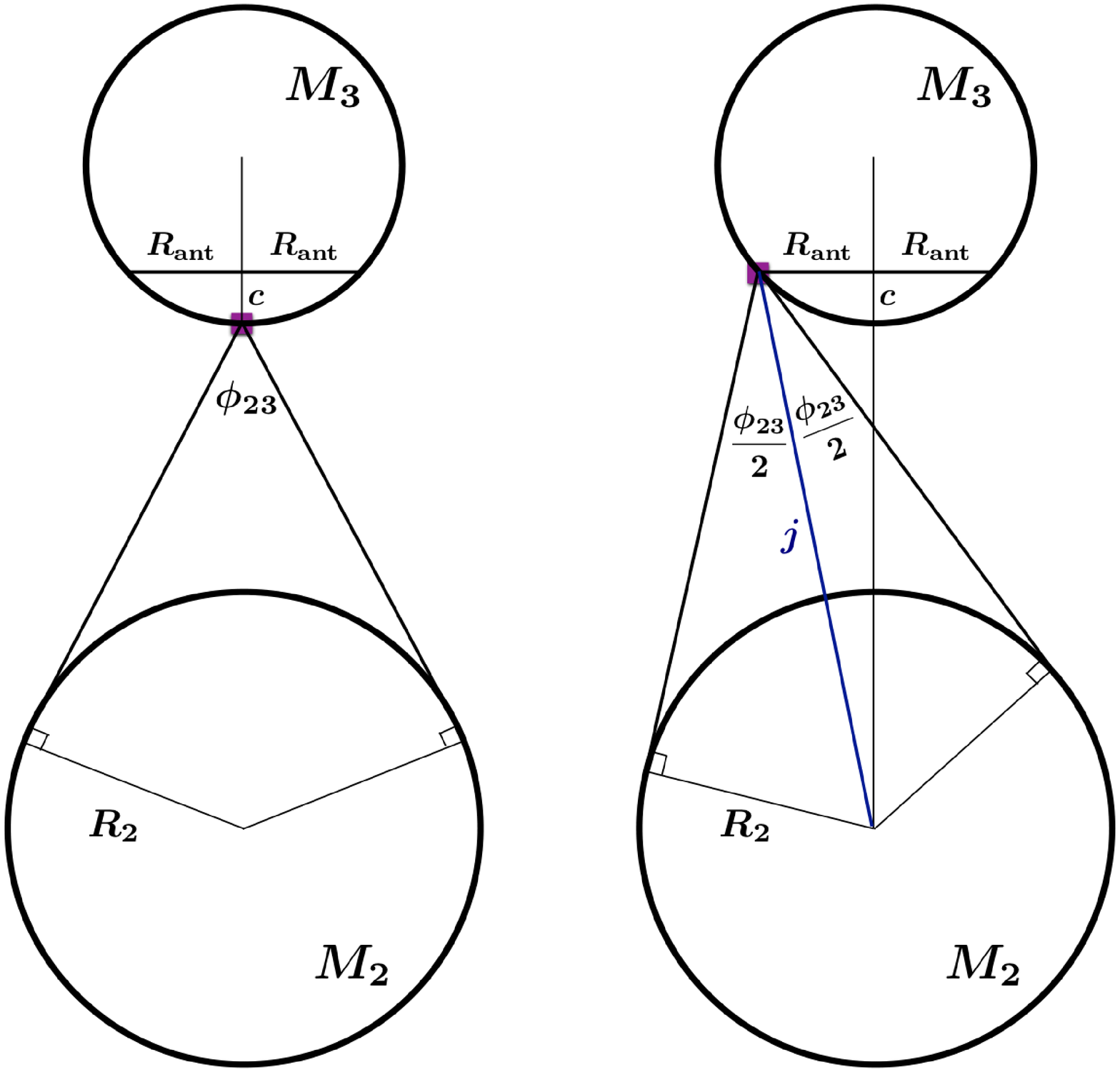}
\caption{
Angular diameter ($\phi_{23}$) of occulter when the viewer (standing on the target $M_3$) 
is co-linear with the syzygy (left-hand panel) 
and when the viewer (square) is at the edge of the umbral or
antumbral shadow (right-hand panel).
These situations respectively maximize and minimize the value of $\phi_{23}$.
Although $R_{\rm ant}$ is used in the diagrams, equally $R_{\rm umb}$ may be adopted.
Equations (\ref{maxphi23}-\ref{jaux}) were derived from the geometry here.
}
\label{AngDiam1}
\end{figure}

\subsection{Angular diameters}

During syzygy, an observer standing on the target in the umbral and antumbral shadow will see some
or none of the radiating primary. Let $\phi_{13}$ represent the angular diameter of the
primary supposing that the occulter does not exist, and let $\phi_{23}$ represent the
angular diameter of the occulter. In diagrams throughout the
paper, we represent the observer as a square.

\subsubsection{Co-linear observers}

For both total and annular eclipses, during syzygy, an observer 
standing on the target anywhere within the shadow will see 
the primary at least partly obscured by the occulter. 
The maximum value of $\phi_{23}$ is achieved when the
observer is co-linear with the syzygy (left panel of
Fig. \ref{AngDiam1}). Then, regardless of the relative size
of the target and occulter,

\[
{\rm max}\left(\phi_{23}\right) = 2 \sin^{-1}\left(\frac{R_2}{r_{23} - R_3} \right),
\ \ \ \ h < r_{13} - n - R_3
\]

\begin{equation}
\ \ \ \ \ \ \ \ \ \ \ \ \ \, 
= 2 \sin^{-1}\left(\frac{R_1}{r_{13} - R_3} \right),
\ \ \ \ h \ge r_{13} - n - R_3
\label{maxphi23}
\end{equation}

\noindent{}such that the upper branch corresponds to an annular eclipse, and the lower branch a total eclipse. The expression in the lower branch is equal to ${\rm max}\left(\phi_{13}\right)$ for both annular and total eclipses.

\subsubsection{Offset observers}

When the observer is not co-linear with the syzygy, but remains
in the umbra or antumbra, then the angular diameters
are smaller. Regardless of the location of the observer within the
shadow, they will see the entire occulter creating an obscuration
on the primary's disc.
For a given architecture, the minimum possible value of 
$\phi_{23}$ is then achieved
when the observer is at the edge of the shadow
(right panel of Fig. \ref{AngDiam1}). Using the helpful auxiliary variables

\begin{equation}
c = R_3 - \sqrt{R_{3}^2 - R_{\rm ant}^2}
,
\end{equation}

\begin{equation}
j = \sqrt{\left(r_{23} - R_3 + c\right)^2 + R_{\rm ant}^2}
.
\label{jaux}
\end{equation}

\noindent{}gives

\[
{\rm min}\left(\phi_{23}\right) =
\]

\[
\ \ \ \ \ \ \ \ \ \ \ \ 2 \sin^{-1}\left[ \frac{R_2}{\sqrt{
\left(r_{23} - \sqrt{R_{3}^2 - R_{\rm ant}^2} \right)^2 
+ R_{\rm ant}^2}} \right],
\]

\[
\ \ \ \ \ \ \ \ \ \ \ \ \ \ \ \ \ \ \ \ \ \ \ \ \ \ \ h < r_{13} - n - R_3;
\]

\

\[
\ \ \ \ \ \ \ \ \ \ \ \ \ 2 \sin^{-1}\left[ \frac{R_1}{\sqrt{
\left(r_{13} - \sqrt{R_{3}^2 - R_{\rm umb}^2} \right)^2 
+ R_{\rm umb}^2}} \right],
\]

\[
\ \ \ \ \ \ \ \ \ \ \ \ \ \ \ \ \ \ \ \ \ \ \ \ \ \ \ \ h \ge r_{13} - n - R_3
\]

\begin{equation}
\end{equation}

\noindent{}such that the upper branch corresponds to an annular eclipse, and the lower branch a total eclipse. Again, the expression in the lower branch is equal to ${\rm min}\left(\phi_{13}\right)$ for both annular and total eclipses. These results hold regardless of whether the
target or occulter is larger.

\subsubsection{Useful benchmark}

The prominence of the eclipse can be measured by 
the fractional area of the primary blocked out
by the occulter

\begin{equation}
g \equiv \left(\frac{\phi_{23}}{\phi_{13}} \right)^2
\label{gval}
\end{equation}

\noindent{}such that a total eclipse corresponds to $g=1$. Here,
$g$ would be equivalent to the so-called ``transit depth'', which
is a useful measure in studies of some exo-planetary systems.

\begin{figure}
\includegraphics[width=8cm]{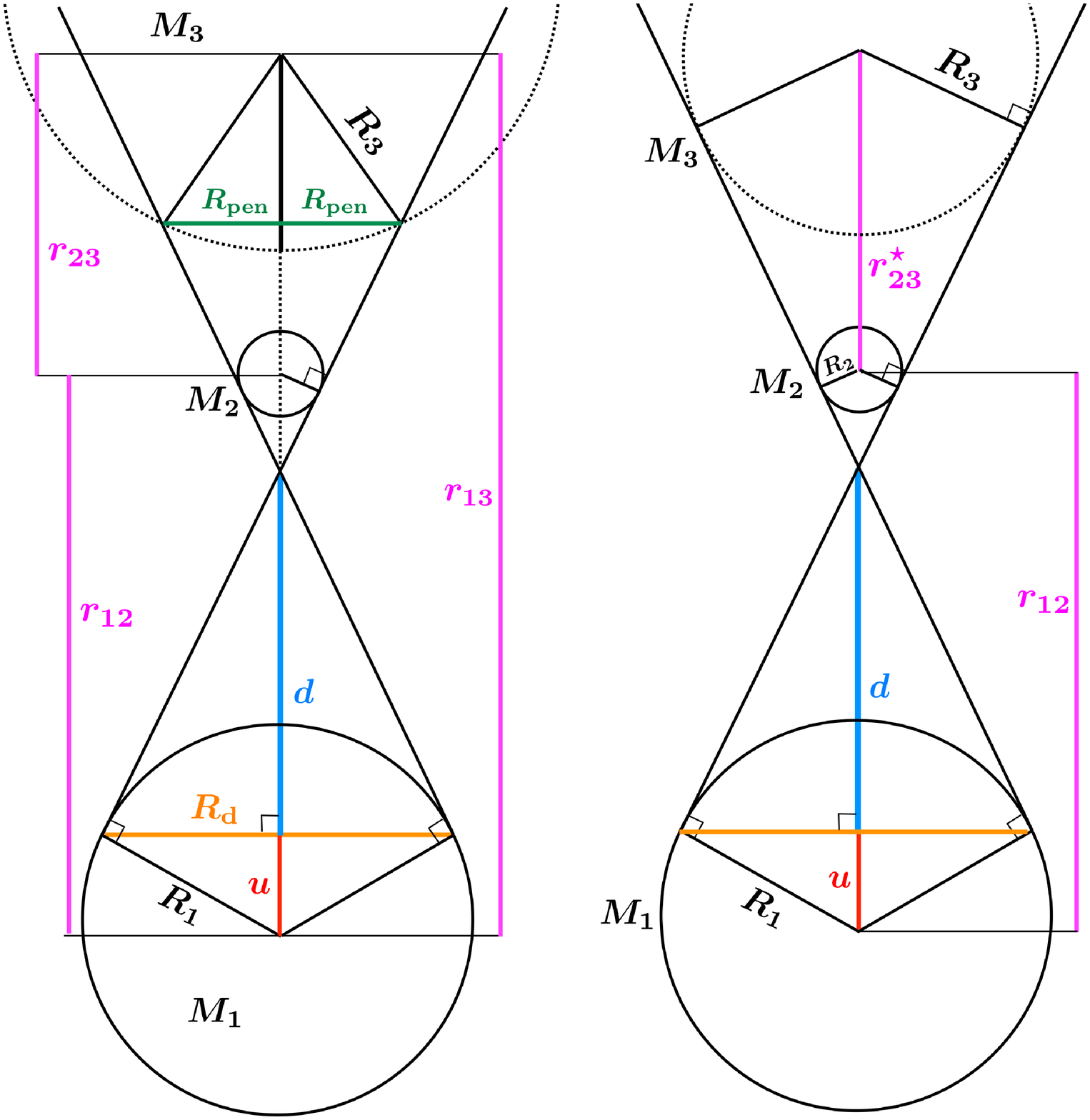}
\caption{
A more detailed geometrical diagram of Fig. \ref{cart2} (left panel)
and the limiting target-occulter distance ($r_{23}^*$) 
for which the entire target is covered in the penumbra (right-hand panel).
For $r_{23} \ge r_{23}^*$, the target is engulfed in the penumbral
shadow. Equations (\ref{dandu}-\ref{R3star}) were derived based on 
the geometry in this diagram.
}
\label{zoomcartPen}
\end{figure}

\section{Partial eclipses}  \label{partial}

If the target is sufficiently large, then regardless if total or annular eclipses occur 
at syzygy, partial eclipses will occur
concurrently for some observers who are outside of the umbral or antumbral shadow
The penumbral shadow is generated from the cone containing the occulter that is drawn 
from internal tangent lines: Figure \ref{cart2} displays a schematic.
We proceed along similar lines in this section as with the last.

Consider the geometry in the left panel of Fig. \ref{zoomcartPen}.
The height of the penumbral
double cone is $d$, and the radius of its base is $R_{\rm d}$.
One obtains

\begin{eqnarray}
u &=& \frac{R_{\rm d}^2}{d}
\label{dandu}
\\
d + u &=& \frac{R_1r_{12}}{R_1 + R_2},
\\
R_{\rm d}^2 &=& \frac{1}{2} \left[d\sqrt{d^2 + 4 R_{1}^2} - d^2\right]
,
\label{Rdcone}
\end{eqnarray}

Comparison with equations (\ref{handn}-\ref{Rcone}) reveals that the
only difference is in the denominator of equation (\ref{handn}).
The simultaneous solution gives

\begin{equation}
d = R_{1} \left[ \frac{r_{12}}{R_1 + R_2} - \frac{R_1 + R_2}{r_{12}} \right]
,
\end{equation}

\begin{equation}
R_{\rm d} = \left(\frac{R_1}{r_{12}}\right)
           \sqrt{r_{12}^2 - \left(R_1 + R_2\right)^2}
,
\end{equation}

\begin{equation}
u = \frac{R_1}{r_{12}} \left(R_1 + R_2\right)
.
\end{equation}

\noindent{}Therefore, $d$, $R_{\rm d}$ and $u$ can be obtained
from $h$, $R_{\rm c}$ and $n$ (equations \ref{handn}-\ref{Rcone}) 
just by substituting $R_2$ with ($-R_2$).

\subsection{Shadow radii}

First, we determine the limiting conditions under which the entire target
is engulfed in the penumbra.

\subsubsection{Total coverage}

This task is aided by the right panel of Fig. \ref{zoomcartPen},
which assumes a static occulter given a fixed value of $r_{12}$.
Let $r_{23}^*$ represent the critical value of $r_{23}$
beyond which the target is engulfed in the penumbral shadow. Then

\begin{equation}
\frac{R_1}{d+u} = \frac{R_2}{r_{12}-d-u} = \frac{R_3}{r_{23}^* + r_{12}-d-u}
\end{equation}

\noindent{}and

\begin{equation}
r_{23}^* = r_{12} \left(\frac{R_3 - R_2}{R_1 + R_2}\right).
\label{r23star}
\end{equation}

\noindent{}Equation (\ref{r23star}) may be compared to equations (\ref{dag1})-(\ref{dag2}),
and shows that if the target is smaller than the occulter, then the target will always
be engulfed. We can also write

\begin{equation}
R_{3}^* = R_2 + \left(R_1 + R_2\right) \frac{r_{23}}{r_{12}}
\label{R3star}
\end{equation}

At $r_{23}^*$ or $R_{3}^*$, less than half of the target
will be bathed in the partial eclipse, similarly to an antumbral
eclipse. The projected radius of the
shadow $R_{\rm pen}$ (see Fig. \ref{cart2}) in this specific case
(right-hand panel of Fig. \ref{zoomcartPen}) is given by

\begin{equation}
\cos{\left(\sin^{-1}\left[\frac{R_{\rm pen}^*}{R_3}\right]\right)} = \frac{R_1}{d+u}
\end{equation}

\noindent{}or

\begin{equation}
R_{\rm pen}^* = \left(\frac{R_3}{r_{12}}\right)
\sqrt{
r_{12}^2 - \left(R_1 + R_2 \right)^2
}
=
\frac{R_{\rm d} R_3}{R_1}
.
\end{equation}

\subsubsection{Some coverage}

When the target is not completely engulfed in the penumbra ($r_{23}^* > r_{23}$),
then in order to determine the shadow radius $R_{\rm pen}$, we again appeal to 
the properties of double cones. See the left panel of Fig. \ref{zoomcartPen}. 
We can make the following substitutions
in equation (\ref{coneeq}): $h \rightarrow d$, $R_{\rm c} \rightarrow R_{\rm d}$,
$\left(x^2 + y^2\right) \rightarrow R_{\rm pen}^2$ and $z \rightarrow r_{13}-u-\sqrt{R_{3}^2-R_{\rm pen}^2}$.
Then, by analogy with the antumbral case,

\begin{equation}
R_{\rm pen} = R_{\rm d} 
\left( 
\frac{r_{13}-u-\sqrt{R_{3}^2-R_{\rm pen}^2}}{d}
- 1
\right)
\end{equation}

\noindent{}or

\[
R_{\rm pen} = \left( \frac{d R_{\rm d}}{d^2 + R_{\rm d}^2} \right)
             \bigg[
                    r_{13}-d-u
\]

\begin{equation}
\ \ \ \ \ \ \ \ \ \ \
-
\sqrt{R_{3}^2
-
\left(\frac{R_{\rm d}}{d}\right)^2
\left[
\left( r_{13}-d-u \right)^2 - R_{3}^2
\right]
}
\bigg]
.
\end{equation}

\begin{figure}
\ \ \ \ \ \ \ \ \ \ \ \ \ \ \ \ \ \ \ \ \
\includegraphics[height=8.0cm]{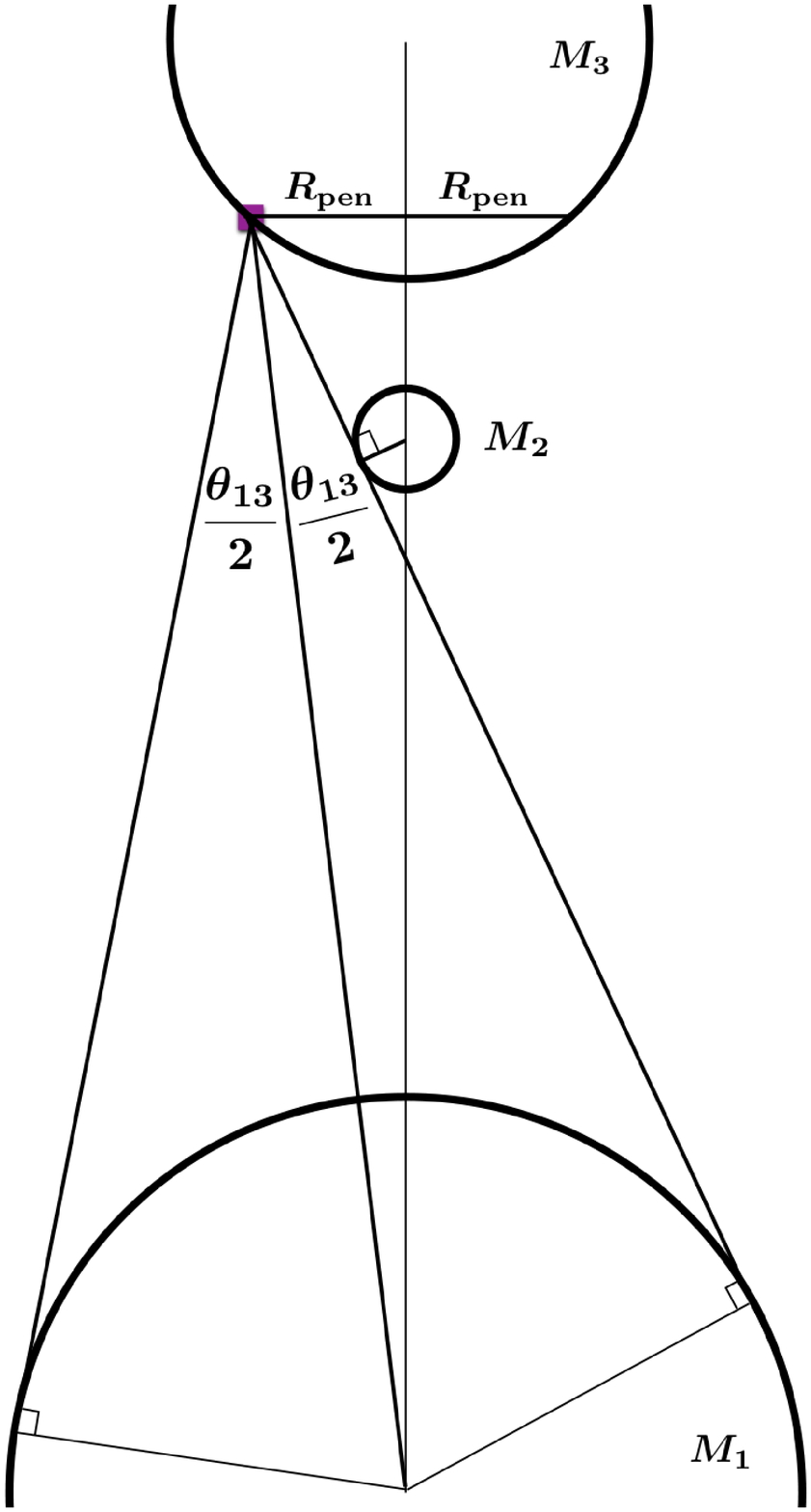}
\caption{
Angular diameter ($\theta_{13}$) of primary when the viewer (square)
is standing at the edge of the penumbral shadow on the target $M_3$.
In this instance, the occulter does not block the primary at all,
giving $\theta_{23} = 0$. 
}
\label{AngDiam3}
\end{figure}

\subsection{Surface area}

The surface area of the target which experiences the partial eclipse is
well-defined as either $\left(S_{\rm pen} - S_{\rm ant}\right)$ or
$\left(S_{\rm pen} - S_{\rm umb}\right)$ provided that $R_{\rm ant}$
or $R_{\rm umb}$ exist and where

\[
S_{\rm pen} = \pi \left[R_{\rm pen}^2 + \left(R_{3} - \sqrt{R_{3}^2 - R_{\rm pen}^2}\right)^2 \right],
\ \ \ \ \ \ \ \ \, r_{23} \le r_{23}^*
\]

\begin{equation}
\end{equation}

\noindent{}We assume that $R_{\rm pen}$ does not exist if $r_{23} > r_{23}^*$.

\subsection{Angular diameters}

\subsubsection{Co-linear observers}

At syzygy, co-linear observers never experience
a partial eclipse. They are instead in the umbra
or antumbra, and so will experience a total eclipse.

\subsubsection{Offset observers at maximum coverage}

Denote the angular diameters of the eclipsed occulter and primary
as seen from an observer (on the target) in the penumbra
as $\theta_{23}$ and $\theta_{12}$. The maximum value of $\theta$
occurs when an observer resides at the boundary 
between the penumbral shadows and the umbral
or antumbral shadows. Hence,

\begin{equation}
{\rm max}\left(\theta_{23}\right) = {\rm min}\left(\phi_{23}\right)
,
\end{equation}

\begin{equation}
{\rm max}\left(\theta_{13}\right) = {\rm min}\left(\phi_{13}\right)
.
\end{equation}

\subsubsection{Offset observers at minimum coverage}

The minimum values of $\theta$ are
achieved at the edge of the shadow. Geometry dictates
that at that location, the target and occulter
appear in their entirety and are grazing.  
Figure~\ref{AngDiam3} illustrates that

\begin{equation}
{\rm min}\left(\theta_{23}\right) = 0
\end{equation}

\noindent{}whereas

\begin{equation}
{\rm min}\left(\theta_{13}\right) =
2 \sin^{-1}\left[ \frac{R_1}{\sqrt{
\left(r_{13} - \sqrt{R_{3}^2 - R_{\rm pen}^2} \right)^2 
+ R_{\rm pen}^2}} \right]
.
\end{equation}


\label{lastpage}
\end{document}